\documentstyle[referee]{laa}

\begin{document}
\thesaurus{07  
	   (07.09.1; 
	   07.13.1;  
	   07.13.2   
	   )}

\title{Meteor Streams and Parent Bodies}

\author{ Jozef Kla\v{c}ka }
\institute{Institute of Astronomy,
Faculty for Mathematics and Physics,
Comenius University, \\
Mlynsk\'{a} dolina,
842 15 Bratislava,
Slovak Republic}
\date{}
\maketitle
\begin{abstract}
Problem of meteor orbit determination for a given parent body is discussed.
Some of the published methods for obtaining meteoroid's orbital elements
at the moment of intersecting Earth's orbit on the basis of
geometrical variation of
parent body's orbital elements are discussed. The main result concerns
the following two facts: i) in real situations physical quantities
for the change of orbital elements
of the parent body must be used, and, ii)
the usage of Southworth and Hawkins (1963) D-criterion yields results not
corresponding to observations.
\end{abstract}

\section{Introduction}
In searching for the relation between  parent body (comet, asteroid) and
its possible meteor stream, one needs to calculate theoretical orbit of a
meteoroid at the time when it may become a meteor. Several methods have been
worked out during the last several tens of years. Their summary may be
found in Svore\v{n} {\it et al.} (1993). Our aim is to discuss some of the
published methods and improve them. We also suggest physical method.

Svore\v{n} {\it et al.} (1993), Neslu\v{s}an {\it et al.} (1994), as examples,
use D-criterion of Southworth and Hawkins (1963) in many of the methods.
Moreover, in final comparison between theoretical and observational
result, this D-criterion is considered to be a decisive quantity.
However, it seems to be not adequate to use this criterion due to
its many ``curious'' properties (Kla\v{c}ka 1995).

We will use standard orbital elements in the paper: $q$ -- perihelion
distance, $e$ -- eccentricity, $i$ -- inclination, $\Omega$ --
longitude of the ascending node, $\omega$ -- argument of perihelion.
Moreover, subscript ``$P$'' refers to parent body, subscript ``$M$'' to
meteoroid which intersects the orbit of the Earth. The Earth's orbit is
supposed to be circular.

\section{General considerations}

If the meteoroid intersects the Earth's orbit, then one node of the
meteoroid's orbit is at distance $r_{E} =$ 1 AU from the Sun. Thus,
we have ($f$ -- true anomaly)
\begin{eqnarray}\label{1}
&~ & \sin \left ( \omega_{M} ~+~ f_{M} \right ) ~=~ 0 ~~~~and~ \nonumber \\
&~ & r_{E} ~=~ \frac{q_{M} \left ( 1 ~+~ e_{M} \right )}{1 ~+~ e_{M}~ \cos f_{M}}  ~.
\end{eqnarray}
Thus, if the ascending node
($\varepsilon > 0 \Rightarrow \sin ( \omega_{M} ~+~ f_{M} ~+~ \varepsilon ) >$
0 -- this corresponds to $\omega_{M} + f_{M} = 2~ \pi$) intersects the orbit of the Earth, then
\begin{equation}\label{2}
1 ~AU \equiv r_{AE} ~=~ \frac{q_{M} \left ( 1 ~+~ e_{M} \right )}{1 ~+~ e_{M}~ \cos \omega_{M}}  ~.
\end{equation}
If the descending node
($\varepsilon > 0 \Rightarrow \sin ( \omega_{M} ~+~ f_{M} ~+~ \varepsilon ) <$
0 -- this corresponds to $\omega_{M} + f_{M} = \pi$) intersects the orbit of the Earth, then
\begin{equation}\label{3}
1~AU \equiv r_{DE} ~=~ \frac{q_{M} \left ( 1 ~+~ e_{M} \right )}{1 ~-~ e_{M}~ \cos \omega_{M}}	~.
\end{equation}

\section{Methods of variation of the orbit in the orbital plane}

Three orbital elements may vary: $q$, $e$, $\omega$. If $\omega_{M} = \omega_{P}$
and $q_{M} = q_{P}$, or, $\omega_{M} = \omega_{P}$ and $e_{M} = e_{P}$, the
situation is trivial -- in both cases only one orbital element ($e$, resp. $q$)
is varied in order to Eq. (2) or Eq. (3) be fulfilled.

\subsection{The argument of perihelion is varied}

Let us suppose that it is sufficient to make a change of $\omega_{P}$, only,
for the purpose of that a meteoroid may intersect the Earth's orbit:
$q_{P} = q_{M}$, $e_{P} = e_{M}$. Eq. (2) yields then
\begin{eqnarray}\label{4}
\omega_{M1} &=& \arccos \left \{ \frac{1}{e_{P}} ~ \left [ ~
		\frac{q_{P} \left ( 1 ~+~ e_{P} \right )}{r_{AE}} ~-~ 1 \right ]
		\right \} ~, \nonumber \\
\omega_{M2} &=& 360^{\circ} ~-~ \omega_{M1} ~.
\end{eqnarray}
Eq. (3) yields
\begin{eqnarray}\label{5}
\omega_{M3} &=& \arccos \left \{ \frac{1}{e_{P}} ~ \left [ ~1 ~-~
		\frac{q_{P} \left ( 1 ~+~ e_{P} \right )}{r_{DE}}  \right ]
		\right \} ~, \nonumber \\
\omega_{M4} &=& 360^{\circ} ~-~ \omega_{M3} ~.
\end{eqnarray}
If $\omega_{M}$ does not exist, then the assumption that the change of only
one element $\omega_{P}$ is sufficient in obtaining meteor, is incorrect.

\subsection{Perihelion distance and eccentricity are simultaneously varied}

Let us consider now that $\omega_{M} = \omega_{P}$ and that $q$ and $e$
may change. If we use also D-criterion of Southworth and Hawkins (1963),
as it is done in Svore\v{n} {\it et al.} (1993) (only $q$ and $e$ are
different for parent body and meteoroid), the minimum simultaneously
change of $q$ and $e$ is given by the condition that the function
\begin{equation}\label{6}
D_{SH}^{2} \equiv ( q_{M} ~-~ q_{P} )^{2} ~+~ ( e_{M} ~-~ e_{P} )^{2}
\end{equation}
gains its minimum value. Besides this condition, Eqs. (2)-(3) also hold
($\omega_{M} = \omega_{P}$). If we substitute $q_{M}$ from Eq. (2) into Eq. (6),
we obtain $D_{SH}$ as a function of $e_{M}$ only and its extremal -- minimum value
can be easily found. The same procedure must be replied for Eqs. (3) and (6).
In both cases it can be analytically proved that only one solution corresponds to
$e_{M} >$ 0. For some parent bodies the minimum value of $e_{M}$ corresponds to
$e_{M} \in (0, 1)$, for the others $e_{M} \in [1, 9/4)$ (the last value is approximate).
In any case, the value of $D_{SH}$ obtains the value, which is even in the interval
$e_{M} \in (0, 1]$ less than it would be if only one of the elements $q_{M}$
and $e_{M}$ is varied.

These analytical results are in contradiction with the
results presented in Svore\v{n} {\it et al.} (1993 -- see Tables 1, 3, 4, 6).

Another important result shows Table 5 in Svore\v{n} {\it et al.} (1993):
the simultaneous optimal change of $q_{M}$ and $e_{M}$ yields value of $D_{SH}$
which is less than the value of $D_{SH}$ if only $q_{M}$ is changing, but
the concidence between theoretical calculations and observed data is better in
the latter case. This result confirms the fact that
D-criterion of Southworth and Hawkins is not good approximation to the real
processes, as it is discussed in Kla\v{c}ka (1995) from the general point of view.

\section{Methods of variation and rotation of the orbit in space}

We will discuss only one method here, in order to stress again
that we do not need any type of D-criterion.

\subsection{Adjustment of the orbit by rotation around the line of apsides}

We will suppose $q_{M} = q_{P}$, $e_{M} = e_{P}$.

Let the coordinate system S' be created in the form that the orbital plane
of the parent body is characterized by the condition $z' =$ 0 and let the
perihelion lies on the positive part of the $x'-$axis: perihelion is
characterized by the unit vector $x_{p}' =$ 1, $y_{p}' = z_{p}' =$ 0.
This unit vector has coordinates $x$, $y$, $z$ in the original coordinate
system S (ecliptical system, plane of the ecliptic: $z =$ 0):
\begin{eqnarray}\label{7}
x_{pP} &=& \cos \Omega_{P} ~ \cos \omega_{P} ~-~  \sin \Omega_{P} ~ \sin \omega_{P}
      ~ \cos i_{P}  \nonumber \\
y_{pP} &=& \sin \Omega_{P} ~ \cos \omega_{P} ~+~  \cos \Omega_{P} ~ \sin \omega_{P}
      ~ \cos i_{P}  \nonumber \\
z_{pP} &=& \sin \omega_{P} ~ \sin i_{P} ~.
\end{eqnarray}
Let the unit vector normal to the orbital plane of the parent body has coordinates
$x_{n}' = y_{n}' =$ 0, $z_{n}' =$ 1. This normal unit vector is characterized by the
following coordinates in the system S:
\begin{eqnarray}\label{8}
x_{nP} &=& \sin \Omega_{P} ~ \sin i_{P}  \nonumber \\
y_{nP} &=& -~ \cos \Omega_{P} ~ \sin i_{P} \nonumber \\
z_{nP} &=& \cos i_{P} ~.
\end{eqnarray}

Now, we make a rotation around the $x'-$axis in an angle $\Phi$. As a result
we obtain a new orbit -- orbital plane of the meteoroid -- characterized by
new normal unit vector, which has coordinates in the system S:
\begin{eqnarray}\label{9}
x_{nM} &=& \sin \Omega_{M} ~ \sin i_{M}  = \nonumber \\
       &=& ( \cos \Omega_{P} ~ \sin \omega_{P} ~+~  \sin \Omega_{P}
	  ~ \cos \omega_{P} ~ \cos i_{P} ) ~ \sin \Phi	~+~  \nonumber \\
       & & +~ ( \sin \Omega_{P} ~ \sin i_{P} ) ~  \cos \Phi  \nonumber \\
y_{nM} &=& -~ \cos \Omega_{M} ~ \sin i_{M} = \nonumber \\
       &=&  ( \sin \Omega_{P} ~ \sin \omega_{P} ~-~  \cos \Omega_{P}
	  ~ \cos \omega_{P} ~ \cos i_{P} ) ~ \sin \Phi	~-~  \nonumber \\
       & & -~ ( \cos \Omega_{P} ~ \sin i_{P} ) ~ \cos \Phi   \nonumber \\
z_{nM} &=& \cos i_{M} = \nonumber \\
       &=& -~ ( \cos \omega_{P} ~ \sin i_{P} ) ~ \sin \Phi ~+~
	   ( \cos i_{P} ) ~ \cos \Phi ~.
\end{eqnarray}

Unit vector of the perihelion of the meteoroid's orbit is
\begin{eqnarray}\label{10}
x_{pM} &=& \cos \Omega_{M} ~ \cos \omega_{M} ~-~  \sin \Omega_{M} ~ \sin \omega_{M}
      ~ \cos i_{M}  \nonumber \\
y_{pM} &=& \sin \Omega_{M} ~ \cos \omega_{M} ~+~  \cos \Omega_{M} ~ \sin \omega_{M}
      ~ \cos i_{M}  \nonumber \\
z_{pM} &=& \sin \omega_{M} ~ \sin i_{M} ~,
\end{eqnarray}
and it is the same unit vector as the unit perihelion vector of the parent body
($x'-$axis -- the line of apsides -- is fixed during the rotation).

Thus, we have one set of three equations given by the equality of the
right-hand-sides of Eqs. (7) and (10) and another set of three equations given
by Eq. (9). For a given angle $\Phi$ Eq. (9) determines quantities
$i_{M}$ and $\Omega_{M}$, and, then, Eqs. (7) and (10) determine the angle
$\omega_{M}$. The angle $\Phi$, for which Eqs. (2) or (3) hold, may be
easily found.

We have found the angle of rotation $\Phi$ around the line of apsids which
determines the orbit of a meteoroid to become a crosser of the Earth's orbit.
Thus, we have found all the required elements: $i_{M}$, $\Omega_{M}$ and
$\omega_{M}$.
We do not need any knowledge about the D-criterion of Southworth and Hawkins
as it is in the method presented in Svore\v{n} {\it et al.} (1993).

\section{Physical Methods}

Let us consider
physical access to the problem.

\subsection{Energy-H$_{z}$-$\pi$ Method}

By this method we consider simple model in which the following quantities of
the meteoroids are equivalent to those of the parent body: semimajor-axis $a$,
z-component of angular momentum $\sqrt{a~(1~-~e^{2})} \cos i$,
$\pi ~=~ \omega ~+~ \Omega$.
Considering Eqs. (2) and (3), the minimization of
the magnitude of the difference of the angular momenta vectors between
meteoroid and its parent body
$| \vec{\bigtriangleup H} |$ $\equiv$
$| \vec{H} ~-~ \vec{H}_{c}|$ (global and local minima) yields
several theoretical radiants -- those with the smallest values of
$| \vec{\bigtriangleup H} |$ should be realized.

%

\subsection{Energy-C$_{1}$-C$_{2}$-C$_{3}$ Method}

By this method we consider the model of Babadzhanov and Obrubov (1987)
in which the four quantities
of the meteoroids are equivalent to those of parent body: semimajor-axis $a$,
z-component of angular momentum $\sqrt{a~(1~-~e^{2})} \cos i$ $\equiv$ $C_{1}$,
$C_{2} = e^{2}~ (0.4 ~-~ (\sin i)^{2} ~ (\sin \omega)^{2})$, $C_{3} \equiv$
$\pi$ $=$ $\omega ~+~ \Omega$. Possible orbits of meteoroid and the corresponding
radiants are found on the basis of these four conservation quantities.
The physical idea of the method yields that all the obtained radiants are
equivalently significant and they all should exist in reality.

\section{Application of the Physical Methods and Method of the Section 4.1}

As an application of the mentioned physical methods we present theoretical
orbital elements and radiants of meteoroids of the comet
73P/Schwassmann-Wachmann 3 for the year 2006. Orbital elements of the
comet, considered in our calculations, are: $q_{c} = 0.93973$ AU,
$e_{c} = 0.69331$, $i_{c} = 11.398^{\circ}$,
$\omega_{c} = 198.824^{\circ}$, $\Omega_{c} = 69.883^{\circ}$; these
elements were obtained by the authors of the paper Gajdo\v{s} {\it et al.}
(1998), where also theoretical radiants by other methods (sections 3 and 4
of this paper) are presented.

The Energy-H$_{z}$-$\pi$ method yields, in principle, four theoretical
radiants. The values of the last column of the Table 1 suggest, however,
that only one radiant should correspond to reality -- that given by the
first row in Table 1.

\begin{table}
\caption{Energy-H$_{z}$-$\pi$ Method.}
\begin{flushleft}
\begin{tabular}{rrrrrrrrr}
\hline
\hline
$q [AU]$ & $e$ & $i [^{\circ}]$ & $\omega [^{\circ}]$ & $\Omega [^{\circ}]$ & $\lambda_{\odot} [^{\circ}]$ & $\alpha [^{\circ}]$
& $\delta [^{\circ}]$ & $| \vec{\bigtriangleup H} |^{2}$ \\
\hline
0.9470 & 0.6910 & 12.3 & 209.5	&  59.2 &  59.2 & 211.4 &    26.9 & 0.0009 \\
0.9176 & 0.7001 &  8.1 & 143.1	& 125.6 & 125.6 & 172.0 &    28.9 & 0.0143 \\
0.8954 & 0.7078 &  0.1 & 318.4	& 310.3 & 130.3 & 163.5 &     6.7 & 0.0205 \\
0.8954 & 0.7078 &  0.1 &  41.6	& 227.1 &  47.1 & 193.9 & $-$ 6.2 & 0.0207 \\
\hline
\hline
\end{tabular}
\end{flushleft}
\end{table}

The Energy-C$_{1}$-C$_{2}$-C$_{3}$ method yields also four radiants. The
physical idea of the method predicts the real existence of all the four radiants
(see Table 2).
(We mention that no meteor stream existed for the year 1930 according to
Energy-C$_{1}$-C$_{2}$-C$_{3}$ method.)



\begin{table}
\caption{Energy-C$_{1}$-C$_{2}$-C$_{3}$ Method.}
\begin{flushleft}
\begin{tabular}{rrrrrrrr}
\hline
\hline
$q [AU]$ & $e$ & $i [^{\circ}]$ & $\omega [^{\circ}]$ & $\Omega [^{\circ}]$ & $\lambda_{\odot} [^{\circ}]$ & $\alpha [^{\circ}]$
& $\delta [^{\circ}]$  \\
\hline
0.9263 & 0.6977 &  9.6 & 214.9	&  53.9 &  53.9 & 206.7 &     18.8 \\
0.9263 & 0.6977 &  9.6 & 145.2	& 123.6 & 123.6 & 173.9 &     33.1 \\
0.9263 & 0.6977 &  9.6 & 325.2	& 303.6 & 123.6 & 151.1 & $-$ 17.9 \\
0.9263 & 0.6977 &  9.6 &  34.9	& 233.9 &  53.9 & 183.7 & $-$ 32.1 \\
\hline
\hline
\end{tabular}
\end{flushleft}
\end{table}

Finally, we present also radiants for the year 2006 obtained by the method
described in section 4.1. Again, also this
method yields, in principle, four theoretical
radiants. The values of the last column of the Table 3 suggest, however,
that only one radiant should correspond to reality -- that given by the
first row in Table 3.
(We mention that no meteor stream existed for the year 1930 according to
this method -- $q_{c} >$ 1.0 AU.)

\begin{table}
\caption{Adjustment of the orbit by rotation around the line of apsides method
-- method decribed in section 4.1}
\begin{flushleft}
\begin{tabular}{rrrrrrrrr}
\hline
\hline
$q [AU]$ & $e$ & $i [^{\circ}]$ & $\omega [^{\circ}]$ & $\Omega [^{\circ}]$ & $\lambda_{\odot} [^{\circ}]$ & $\alpha [^{\circ}]$
& $\delta [^{\circ}]$ & $\Phi [^{\circ}]$ \\
\hline
0.9397 & 0.6933 &   7.0 & 211.5  &  57.1 &  57.1 & 203.1 &     14.0 &	4.8 \\
0.9397 & 0.6933 &   7.0 & 328.5  & 299.7 & 119.7 & 154.0 & $-$ 12.9 &  16.8 \\
0.9397 & 0.6933 & 173.0 & 328.5  & 237.1 &  57.1 & 337.7 & $-$ 13.5 & 184.8 \\
0.9397 & 0.6933 & 173.0 & 211.5  & 119.7 & 119.7 &  19.2 &     12.3 & 196.8 \\
\hline
\hline
\end{tabular}
\end{flushleft}
\end{table}

The first two rows of Table 2 correspond to the first two rows of Table 1.
The third row of Table 2 corresponds to the second row of Table 3.
However, the last columns of Tables 1 and 3 show that only the first row
of Tables 1 and 3 can correspond to the reality. In other words, the real
existence of all the four radiants given by the
Energy-C$_{1}$-C$_{2}$-C$_{3}$ method seems to be improbable.

\section{Discussion}

We have discussed some of the published methods
of calculating meteoroid
orbits crossing the Earth's orbit if the parent body's orbit is known.

If the parent body's orbit is changed in so a trivial geometric way as it
is often done,
one should try to avoid to any further requirement,
e. g., to use D-criterion of Southworth and Hawkins. The reason is that
this criterion yields nonphysical results, as it is pointed out
in the last paragraph of the section 3.2 and discussed in Kla\v{c}ka (1995).

The use of simple methods has one great advantage -- calculations are very
simple. However, since changes of orbital elements in such simple methods are
only of geometrical character, the methods do not correspond to any real
physical process --
the methods do not use any physical basis. Since real physics is very
complicated and we do not know exact perturbations on meteoroids (the
importance of nongravitational effects), we have to make predictions for
a new possible pairs ``parent body -- meteor stream''
on the basis of a simple physics, or,
on the basis of the
known pairs.


Possible way is to find a function $F$ (containing
five orbital elements in an independent way) which is a good
approximation to reality and does not exhibit any inconsistencies.
Suggestion of this type may be found in Kla\v{c}ka (1995), where also
general method is presented -- special case of this method is
Energy-H$_{z}$-$\pi$ method. The method in reality corresponds to the
following set of equations:
\begin{equation}\label{11}
\beta_{iM} ~=~ f_{i} ( \beta_{jP} ) ~, ~~~ i, j = 1 ~to~5~,
\end{equation}
where $\beta-$s represent orbital elements ($P$ -- parent body, $M$ --
meteoroid, meteor stream) and Eqs. (2) or (3) hold.  The unknown functions
must be found (approximated) on the basis of observational data -- pairs
``parent body -- meteor stream''. The conditions $f_{i} = \beta_{iP}$
($i =$ 1 to 5) hold if $\beta_{jP}$ fulfill Eqs. (2) or (3). Thus, the
method presented in Kla\v{c}ka (1995) is equivalent to the following sets
of equations:
\begin{eqnarray}\label{12}
\beta_{iM} &=& \beta_{iP} ~+~ \left \{
      \frac{q_{P} \left ( 1 ~+~ e_{P} \right )}{1 ~+~ e_{P}~ \cos \omega_{P}}
      ~-~ 1 \right \} ~f_{1i} ( \beta_{jP} ) ~, ~~~or \nonumber \\
\beta_{iM} &=& \beta_{iP} ~+~ \left \{
      \frac{q_{P} \left ( 1 ~+~ e_{P} \right )}{1 ~-~ e_{P}~ \cos \omega_{P}}
      ~-~ 1 \right \} ~f_{2i} ( \beta_{jP} ) ~, \nonumber \\
& & ~~~ i, j = 1 ~to~5~,
\end{eqnarray}
where $f_{1i}$ and $f_{2i}$ are finite, and, moreover, Eqs. (2) or (3) hold.

In any case, one must be aware of the fact that real meteoroid streams
have various dispersions of orbital elements and that our methods of
calculating new pairs ``parent body -- meteor stream'' are only a first
approximation.

{\bf In principle, \underline{no reasonable metric} for various trajectories
of parent bodies and their meteoroids exists. Gravitational and nongravitational
forces acting on parent bodies and their meteoroids are different in various
parts of orbital element's phase space, and, no global metric can be constructed
on the basis of the know pairs ``parent body -- meteor stream''. Useful methods
for predicting new pairs ``parent body -- meteor stream'' must be based on the
known pairs ``parent body -- meteor stream'' and inevitable weighting factors,
functions of parent body's orbital elements, present in ``metric'' disturb
the basic property of metric -- its symmetricity.}

It is hard to make reasonable predictions for meteor streams corresponding to
new parent bodies lying outside the zone of the known parent bodies for which
we know the relation ``parent body -- meteor stream''. Only extrapolations can
be made in this case. Much better predictions are obtained for the inner part
of the zone of the phase space. Energy-H$_{z}$-$\pi$ method seems to yield
good results in any case.

\section{Appendix}
Derivation of equations of section 4.1 is presented here, on the request
of the referee.

Let the Cartesian coordinate system S' is created from the Cartesian coordinate
system S by the rotation characterized with Eulerian angles $\Omega$,
$i$ and $\omega$. If $x$, $y$, $z$ are coordinates of a vector in the system
S, then $x'$, $y'$, $z'$ are coordinates of the same vector in the system S'.
We have

\[ \left ( \begin{array}{c}
x \\ y \\ z
\end{array} \right )
=
\left ( \begin{array}{ccc}
C_{11} ~~ & ~~~ C_{12} ~~ & ~~~C_{13} \\
C_{21} ~~ & ~~~ C_{22} ~~ & ~~~C_{23} \\
C_{31} ~~ & ~~~ C_{32} ~~ & ~~~C_{33} \\
\end{array} \right )
\left ( \begin{array}{c}
x' \\ y' \\ z'
\end{array} \right )	~, \]
\noindent
and the transformation matrix $C$ is of the form
\[ \left ( \begin{array}{ccc}
\cos \Omega ~\cos \omega - \sin \Omega ~\sin \omega ~\cos i ~~ &
~~~- \cos \Omega ~\sin \omega - \sin \Omega ~\cos \omega ~\cos i ~~ &
~~~\sin \Omega ~\sin i \\
\sin \Omega ~\cos \omega + \cos \Omega ~\sin \omega ~\cos i ~~ &
~~~- \sin \Omega ~\sin \omega + \cos \Omega ~\cos \omega ~\cos i ~~ &
~~~- \cos \Omega ~\sin i \\
\sin \omega ~\sin i & ~~~\cos \omega ~\sin i & ~~~\cos i
\end{array} \right ) \]
\noindent
(Transformation is orthogonal and so inverse transformation can be easily
obtained, using transposed matrix.)

Unit vectors of perihelia are defined by conditions $x' =$ 1, $y' = z' =$ 0,
and so the last equation immediately yields Eqs. (7) and (10).
Unit vector normal to the orbital plane is defined by conditions $x' = y' =$ 0,
$z' =$ 1, and, again,
the last equation immediately yields Eq. (8).

Moreover, if we make a rotation characterized by an angle $\Phi$ around
$x'-$axis, we obtain
\[ \begin{array}{lllll}
x'' &= & x' & & \\
y'' &= & y' ~\cos \Phi ~ &+~ &z' ~\sin \Phi \\
z'' &= -~ & y' ~\sin \Phi ~ &+~ &z' ~\cos \Phi ~.
\end{array}  \]
\noindent
Now, if we put $x'' = y'' =$ 0, $z'' =$ 1, then
$x' =$ 0, $y' = -~ \sin \Phi$, $z' = \cos \Phi$, and the first equation
of the appendix yields results which are consistent with Eq. (9).

\acknowledgements
Special thanks to the firm ``Pr\'{\i}strojov\'{a} technika, spol. s r. o.''.
This work was partially supported by Grants VEGA No. 1/4304/97 and
1/4303/97.
The author wants to thank to the Organizing comittee for the possibility
of presenting the greater part of this paper at the conference
ACM'96. Small comments of referees of both papers (the first -- Meteor
stream membership criteria) are also acknowledged, especially that of
T. Jopek which enabled to remove one incorrect sign (1995).

\end{document}